\pdfoutput=1
\documentclass[aps,pra,twocolumn,superscriptaddress,showpacs]{revtex4}
\usepackage{graphicx}
\usepackage{setspace}
\usepackage{titlesec}
\usepackage{epsfig}
\usepackage{amssymb}
\usepackage{amsmath}
\usepackage{epstopdf}

\begin{document}

\title{Magnetic confinement of neutral atoms based on patterned vortex 
distributions in superconducting disks and rings}

\author{B.~Zhang}
\affiliation{Division of Physics and Applied Physics, Nanyang Technological 
University,  21 Nanyang Link, Singapore 637371, Singapore}
\affiliation{Centre for Quantum Technologies, National University of 
Singapore, 3 Science Drive 2, Singapore 117543, Singapore}
\author{M.~Siercke}
\affiliation{Division of Physics and Applied Physics, Nanyang Technological 
University,  21 Nanyang Link, Singapore 637371, Singapore}
\affiliation{Centre for Quantum Technologies, National University of 
Singapore, 3 Science Drive 2, Singapore 117543, Singapore}
\author{K.S.~Chan}
\affiliation{Division of Physics and Applied Physics, Nanyang Technological 
University,  21 Nanyang Link, Singapore 637371, Singapore}
\author{M.~Beian}
\affiliation{Division of Physics and Applied Physics, Nanyang Technological 
University,  21 Nanyang Link, Singapore 637371, Singapore}
\affiliation{Centre for Quantum Technologies, National University of 
Singapore, 3 Science Drive 2, Singapore 117543, Singapore}
\author{M.J.~Lim}
\affiliation{Department of Physics, Rowan University, 201 Mullica Hill 
Road, Glassboro, NJ 08028, USA}

\author{R.~Dumke}
\email{rdumke@ntu.edu.sg}
\affiliation{Division of Physics and Applied Physics, Nanyang Technological 
University,  21 Nanyang Link, Singapore 637371, Singapore}
\affiliation{Centre for Quantum Technologies, National University of 
Singapore, 3 Science Drive 2, Singapore 117543, Singapore}

\date{\today}

\date{\today}

\begin{abstract}
We propose and analyze neutral atom traps generated by vortices imprinted 
by magnetic field pulse sequences in type-II superconducting disks and rings. 
We compute the supercurrent distribution and magnetic field resulting from 
the vortices in the superconductor. Different patterns of vortices can be 
written by versatile loading field sequences. We discuss in detail procedures 
to generate quadrupole traps,  self-sufficient traps and ring traps based on 
superconducting disks and rings. The ease of creating these traps and the low 
current noise in supercurrent carrying structures makes our approach 
attractive for designing atom chip interferometers and probes.   
\end{abstract}


\maketitle 

\section{Introduction}

Recently superconducting atom chips have generated a lot of interest due 
to their attractive properties, such as the Meissner effect for type-I 
superconductors and vortices for type-II superconductors
 \cite{Cano,Cano08,Emmert09}. Cold atoms have been trapped and manipulated 
 near superconducting surfaces \cite{Mukai,Nirrengarten,Emmert09} and 
 Bose-Einstein condensation (BEC) on a superconducting atom chip has been 
 reported recently \cite{Roux}. Thermal and technical noise in proximity 
 to superconducting surfaces have been shown both theoretically and 
 experimentally to be significantly reduced compared to conventional atom 
 chips 
 \cite{Scheel05a,Skagerstam06a,Hohenester07, Nogues, Hufnagel, 
 Kasch,Emmert, Fermani09}. Superconducting atom chips have the potential 
 to coherently interface atomic and molecular quantum systems with quantum 
 solid state devices \cite{Tian,Andre,Tordrup,Verdu,Sorensen}.
 Furthermore, ultracold atoms can be used to probe properties of 
 superconductors, such as the dynamics of vortices in type-II superconductors 
 \cite{Scheel07, Shimizu}.

Previously we reported on the design and realization of several types of 
magnetic traps involving linear type-II superconducting microstructures 
\cite{Zhang10,Mueller09, Mueller10}. These traps are formed by the magnetic 
field carried by vortices imprinted in the superconductor to confine ultracold 
atoms. Loading of the vortices is accomplished by tailored external magnetic 
pulse sequences applied to the superconductor in the mixed state. In a 
different approach vortex-based magnetic traps have been observed after 
cooling a superconducting disk through the critical temperature in the 
presence of an external magnetic field \cite{Shimizu}.

In this article, we extend our investigation of vortex based microtrap 
geometries to type-II superconducting disks and rings. To this end we have 
developed a numerical approach to calculate the vortex patterns imprinted 
after applying various external magnetic field pulses to the superconducting 
structure. Using this numerical approach we 
design quadrupole traps, self-sufficient traps and ring traps in the various 
superconducting geometries. We demonstrate that, by simple application of an 
additional bias field, a single quadrupole trap may be deformed into a ring. 
Considering the low current noise expected from the persistent supercurrents 
generating these traps, and the low thermal noise expected from 
superconducting atom chips, this method may be an attractive way to generate 
low-noise on-chip atom interferometry. Compared to other low noise traps 
generated by permanent magnets our system retains one desirable feature: The 
ability to control the trap parameters and geometry from shot to shot.

The paper is organized as follows. Sec. \ref{sec:theoy} describes the 
equations of motion for the sheet current density $J$ in a type-II 
superconducting strip, disk and ring in an external magnetic field. To 
validate our numerical approach we compute the supercurrents in a
superconducting strip, and compare them with analytical results \cite{Zhang10}. For the two-dimensional structures considered in this paper no explicit analytical solution is known. 
In Sec. \ref{sec:result} we design several types of traps for cold atoms by 
tailoring the magnetic field pulses which induce vortex patterns in the 
superconducting disk and ring. Sec. \ref{sec:con} summarizes our results.

\section{Loading vortices}
\label{sec:theoy}

\begin{figure}[!h]
 \includegraphics[width=0.5\textwidth]{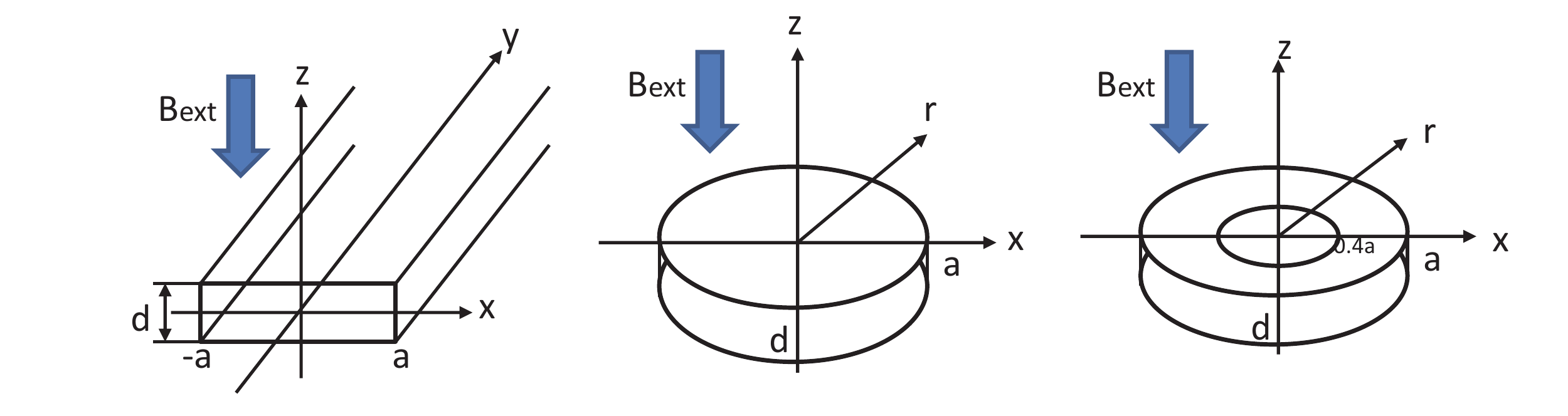}
   \caption{\label{fig:sketch} (Color online) Schematic of the investigated type-II 
   superconductor geometries. All geometries are assumed to be in the thin 
   film limit with a thickness d. Shown are a strip of width $2a$ (left), a 
   disk with radius $a$ (center) and a ring with an outer radius of $a$ and an 
   inner radius of $0.4a$ (right). The magnetic field pulses are applied 
   perpendicular to the surface.}
 \end{figure}
 Consider a type-II thin superconducting strip, disk and ring in a magnetic 
 field perpendicular to their surface ($xy$ plane) as shown in 
 fig.\ref{fig:sketch}. Here $d$ denotes the thickness of the superconductors 
 and $a$ denotes the half width of the strip and the (outer) radius of the 
 disk and ring. The strip is assumed to be infinitely long along $y$, so that 
 we can restrict our simulation to the $xz$ plane. Similarly, we can reduce 
 our simulation to the $rz$ plane for the disk and ring due to rotational 
 symmetry. Throughout the paper we will assume that the temperature of the 
 superconductors is below the critical value $T_c$ and that the thickness 
 $d$ is small compared to the other length scales of the system. When the 
 external field  $B_{ext}$ exceeds the lower critical field $B_{c1}$, but 
 lies below the upper critical value $B_{c2}$, magnetic flux  starts to 
 penetrate the superconductor in the form of Abrikosov vortices 
 \cite{Abrikosov}. 
 To facilitate our numerical analysis we model these 
 vortices together with the screening current as a macroscopic supercurrent $J(u,t)$. 
 Our approach closely follows the 
 method outlined in \cite{Brandt94,Brandt95} which makes use of the continuum approximation where the flux line spacing and the magnetic field penetration depth are small compared to all other relevant length scales considered \cite{Brandt99PRB59}. From the distribution of the 
 supercurrent we then compute the total magnetic field outside of the 
 superconductor using the Biot-Savart law.
 We approximate the current density in the superconductor to be 
 constant over its thickness $d$, which is valid in the thin-film approximation where $d$ is smaller than the London penetration depth of the superconductor. The sheet current density is then defined as 
 $J(x)=j(x)d$, where $j(x)$ is the local current density.
    
The perpendicular field component $B_z(x)$ generated in the specimen plane 
$z=0$ by a constant $B_{ext}$ and by the sheet current $J(x)$ along $y$ for 
the strip can be written as \cite{Brandt94}:
\begin{equation}
 B_z(x)  =   B_{ext}+\frac{\mu _0}{2\pi} \int^a_0 J(u) \Big( \frac{1}{x-u}-
 \frac{1}{x+u} \Big ){\rm d}u \, ,
\end{equation}
For cylindrically symmetric structures, where $J(r)$ is circling clockwise, 
we can write the perpendicular magnetic field component with $r_0=0$ for disks 
and $r_0=0.4a$ for rings as \cite{Brandt94}: 
\begin{equation}
 B_z(r)  =   B_{ext}+\frac{\mu _0}{2\pi} \int^a_{r_0} J(u) \Big( \frac{E(k)}
 {r-u}-\frac{K(k)}{r+u} \Big ) {\rm d}u \, ,
\label{eq:Bz}
\end{equation}
where $E(k)$ and $K(k)$ are the first and second kind of the complete elliptic 
integrals with $k= \sqrt{4ru}/(r+u)$ and $\mu_0$ is the vacuum permeability 
\cite{Brandt94,Brandt95}. In order to find an expression for the sheet current density $J(x)$ we take into account the time-dependence of $B_{ext}$ 
and use Maxwell's equations, to arrive at the equation of motion for the 
sheet current density $J(x,t)$ of the strip \cite{Brandt94}:
\begin{equation}
 J(x,t) =  \tau \left[ 2\pi x \frac{{\rm d} B_{ext} }{{\rm d }t}+ \int^a_0 
 \frac {\partial J(u,t)}{\partial t} M(x,u) {\rm d}u \right]\, ,
 \label{eq:Jtstrip}
\end{equation}
and $J(r,t)$ for disks and rings \cite{Brandt94}:
\begin{equation}
  J(r,t)  =   \tau \left[ \pi r \frac{{\rm d} B_{ext} }{{\rm d }t}+ 
  \int^a_{r_0} \frac {\partial J(u,t)}{\partial t} Q(r,u) {\rm d}u \right] 
  \, ,
\label{eq:Jtdisk}
\end{equation}
where 
\begin{eqnarray}
M(x,u) & = & \ln \Big |\frac{x-u}{x+u}\Big | \, , \nonumber\\
Q(r,u) & = & -\frac{u}{r} \int _{r_0 / u}^{r/u} \Bigg
(\frac{E(\frac{\sqrt{4\omega}}{1+\omega})}{1-\omega}+
\frac{K(\frac{\sqrt{4\omega}}{1+\omega})}{1+\omega}\Bigg) \omega 
{\rm d} \omega \, , \nonumber \\
\tau & = & \frac{\mu_0 a d }{2 \pi \rho(J) } \, .
\end{eqnarray}
Here, $\rho (J) = \rho _c (J/J_c)^{n-1}$ is the nonlinear resistivity with  
critical value $\rho_c$, and $n=U_0/k_b T$ where $U_0$ is the characteristic activation energy of the superconductor \cite{blatter}. $J_c$ is the critical sheet current 
density \cite{Brandt94,Brandt95}, defined as the maximum current density the superconductor can carry without a transition into the normal state. For the scope of this paper we have chosen the value of $n=19$. This value gives a moderate response of the superconductor to applied fields. The exact value for $n$ is material dependent and does not qualitatively alter the characteristic current distributions discussed for the experimentally applicable values.

We approximate the integration in Eq.\ref{eq:Jtstrip} and \ref{eq:Jtdisk}
 by a Riemann Sum. First, we discretise the width of the strip and disk (or 
 ring) into N elements with the same size $\Delta x = a/N$ or 
 $\Delta r = a/N$($\Delta r = a-(r_0)/N$), and denote the middle point of 
 each element by $x_{n}$ or $r_{n}$, $n=1,2...N$. Then we rewrite  
 Eq.\ref{eq:Jtstrip} in the form of a Riemann Sum in matrix form  
 \begin{equation}
  J(x_{i},t)  =   \tau \left[ 2\pi x_{i} \frac{{\rm d} B_{ext} }{{\rm d }t}
  + \sum ^{j=N}_{j=1} M(x_{i},x_{j}) \frac {\partial J(x_{j},t)}{\partial t}  
  \right] \, .
           \label{eq:JtNstrip}
\end{equation}
To solve the differential equation, we invert 
Eq.\ref{eq:JtNstrip} to find an equation for $\frac {\partial J(x_{j},t)}{\partial t}$
\begin{equation}
  \frac {\partial J(x_{i},t)}{\partial t}  = \sum ^{j=N}_{j=1} M^{-1} 
  (x_{i},x_{j}) \left[ \frac{J(x_{j},t)}{\tau} -  2\pi x_{j} \frac{{\rm d} 
  B_{ext} }{{\rm d }t}  \right] \, ,
           \label{eq:JtN}
\end{equation}
 where $M^{-1} (x_{i},x_{j})$ is the inverse of the matrix $M (x_{i},x_{j})$.
  Now, we can use the Euler method and the initial conditions $J(x_{i},t_0)=0$ and 
  $B_{ext}(t_0) =0$ to compute the sheet current density $J(x_{i},t_n)$ at 
  $t=t_n$ by  
  \begin{align}
   J(x_{i},t_n) & = \Bigg( \sum ^{j=N}_{j=1} M^{-1} (x_{i},x_{j}) \left[ 
   \frac{J(x_{j},t)}{\tau} -  2\pi x_{j} \frac{{\rm d} B_{ext} }{{\rm d }t}  
   \right] \nonumber \\
   & + J(x_{i},t_{n-1}) \Bigg) \times \Delta t     \, ,
           \label{eq:N}
\end{align}
where $\Delta t= t_{n}- t_{n-1}$. The same approach can be used for Eq.\ref{eq:Jtdisk}.
  From the above Eq.\ref{eq:N}, we can see that the final sheet current 
  density $J(x_{i},t_n)$ is 
  only dependent on the initial conditions and $t_n$, not on ${\rm d} B_{ext} /{\rm d }t$. After we obtain 
   $J(x_{i},t_n)$, we compute the $B$ field outside of the superconducting 
   film ($z>0$) using the Biot-Savart law.

\begin{figure}[h]
 \includegraphics[width=0.45\textwidth]{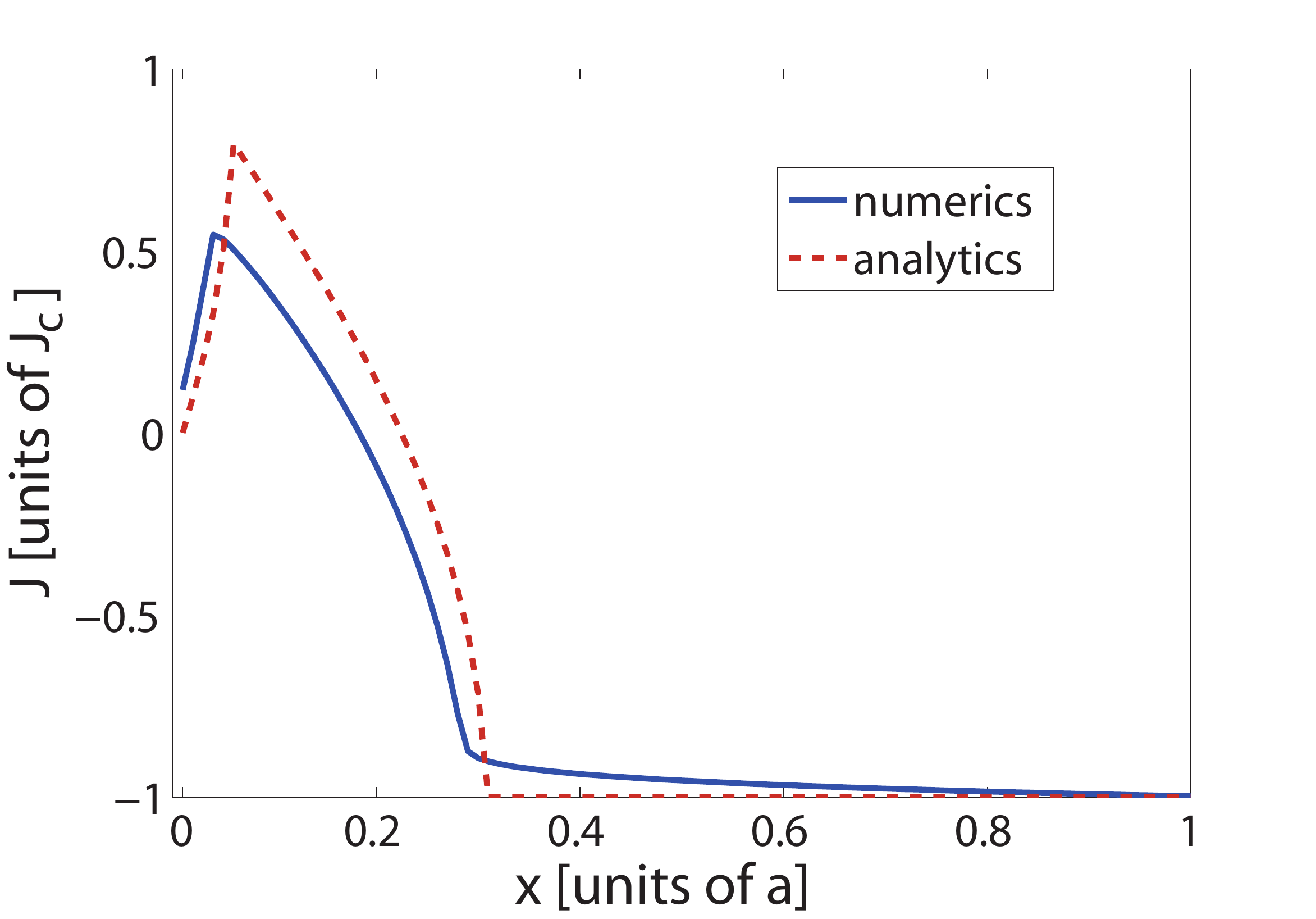}
   \caption{\label{fig:s} (Color online) Distribution of the sheet current density across the 
   strip for $B_{ext}=0 \to \mu_0 J_c \to 0$. Solid line: sheet current 
   density obtained by Eq.\ref{eq:Jtstrip}; Dashed line: sheet current density 
   our former results \cite{Zhang10}. }
 \end{figure}

In order to validate our numerical model, we consider the case of vortices 
induced in a superconducting strip, for which an analytical solution exists \cite{Zhang10}. Applying a magnetic field pulse with a 
magnitude of $B_{ext}$ perpendicular to the strip surface, the magnetic flux 
penetrates the strip. We solve Eq.\ref{eq:Jtstrip} numerically and obtain the 
distribution of supercurrents in the strip plotted in Fig.\ref{fig:s}
(solid lines). The small difference in the curves arises from the fact that the analytical solution assumes that at the edge, where the magnetic flux 
penetrates, the sheet current density takes the critical value. For the 
numerical computation the sheet current density is completely determined by 
Eq.\ref{eq:Jtstrip}.

\section{Magnetic traps}
\label{sec:result}
 
In this section, we apply Eq.\ref{eq:Jtdisk} to compute the sheet current 
density induced in the superconducting disk and ring by various loading 
fields. From the distribution of the sheet current density we then compute the 
magnetic field carried by the vortices by means of the Biot-Savart law. An intuitive picture of the vortex loading process can be found in \cite{Bean90}.

\subsection{Traps generated by a superconducting disk}
\label{sec:disk}

\begin{figure}[!h]
 \includegraphics[width=0.45\textwidth]{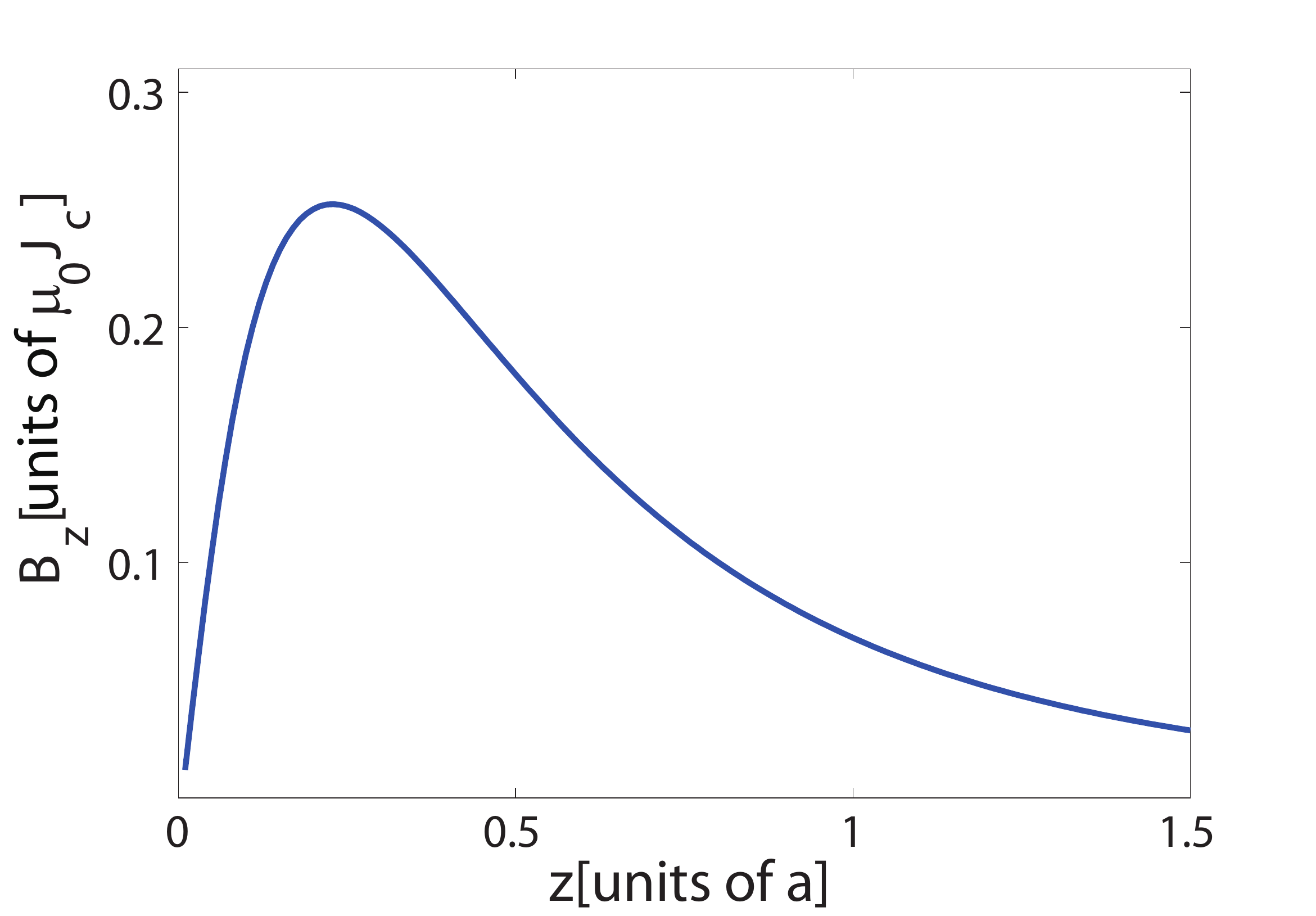}
   \caption{\label{fig:diskB} (Color online) Vertical component $B_z(r,z)$ of the magnetic 
   field carried by vortices loaded by $B_{ext}=0 \to 1.2 \mu_0 J_c \to 0 $ 
   vs $z$ above the disk center $r=0$.}
 \end{figure}

\begin{figure}[!hhh]
 \includegraphics[width=0.45\textwidth]{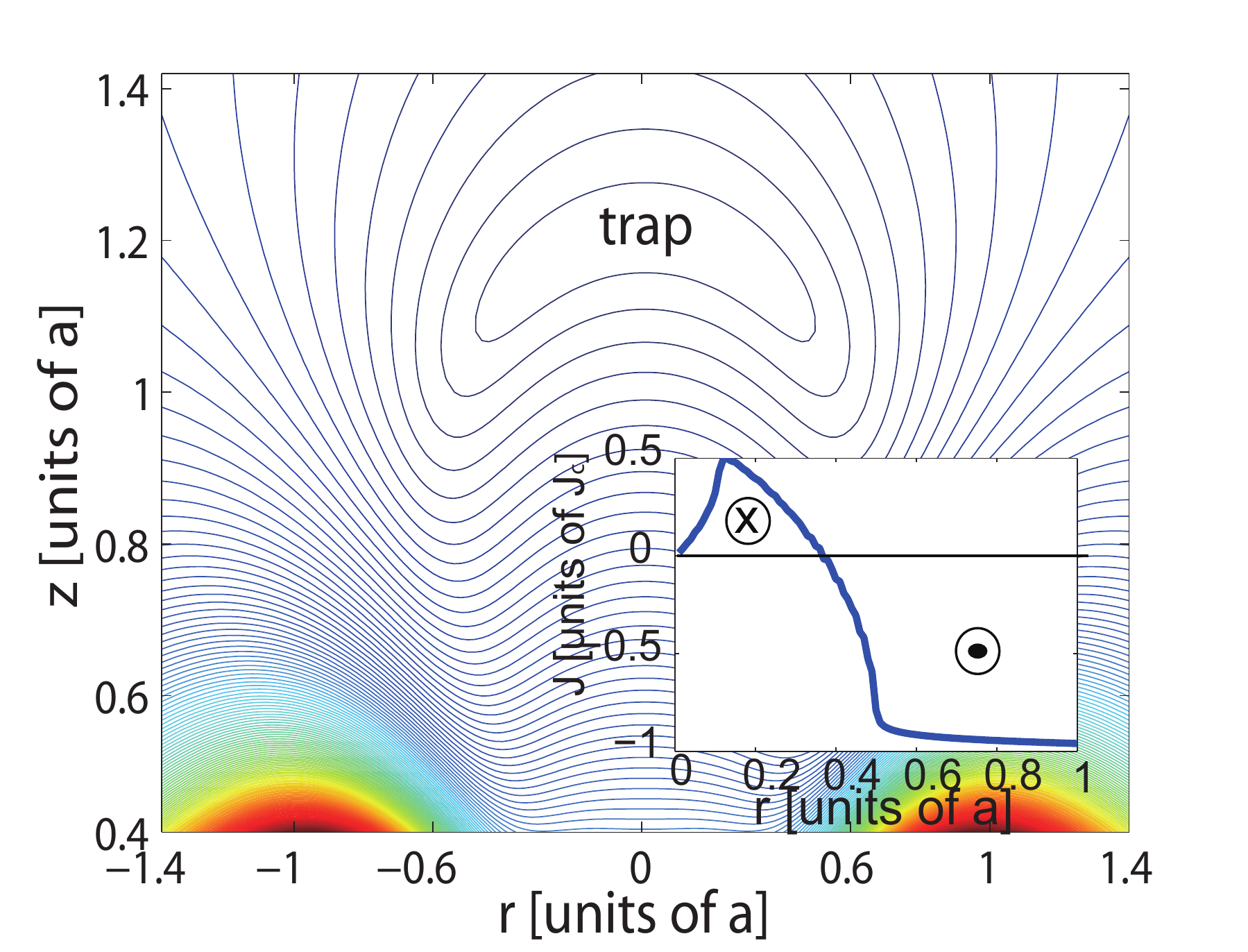}
    \caption{\label{fig:quatrap} (Color online) Equipotential lines showing a quadrupole 
    trap above the disk center formed by the vortex field together with a 
    perpendicular field $B_{bias}=-0.05 \mu_0 J_c$. The vortex loading field 
    pulse has an amplitude of $1.2 \mu_0 J_c$. The field variation per
    contour line is $5\times10^{-3} \mu_0 J_c$. Inset: Distribution of the 
    sheet current density along the radial direction resulting from the 
    loading field pulse.}
 \end{figure}

Consider a disk shaped superconductor in the remanent 
state  with a magnetic loading field pulse of  $B_{ext}=0 \to 1.2 \mu_0 J_c 
\to 0 $. The second critical field in our calculations is approximately $1.8 \mu_0 J_c$.
For symmetry reasons the radial component of the total external field is 
always zero $B_r(r=0,z)=0$ above the disk center.   The $z$ component of the 
total external field $B_z(r=0,z)$ above the disk center at different heights 
is plotted in Fig.\ref{fig:diskB}. $B_z(r,z)$ is always zero at the surface 
center of the disk $(r=0,z=0)$, because the superconductor prevents any 
magnetic field perpendicular to its surface at this point. The inset of 
Fig.\ref{fig:quatrap} shows the supercurrent distribution obtained from 
Eq.\ref{eq:Jtdisk}. We see that the supercurrent in the disk is separated into 
two regions; one with current propagating clockwise and one with current in 
the counter-clockwise direction. Applying a small bias field on the order of 
$B_{ext}/10$ perpendicular to the disk, allows us to cancel $B_z(r=0,z=z_0)$ 
at $z_0$ and create a quadrupole trap for low-field seeking atoms. 
Fig.\ref{fig:quatrap} shows such a trap for $B_{bias}=-0.05 \mu_0 J_c$ and 
$z_0 \sim 1.2a$. Increasing the bias field will move this trap closer to the 
disk surface, however in doing so the bias field will also induce currents 
into the disk, ensuring that the field at the disk surface center remains 0. 
This puts a $J_c$ dependent limit on the trap depth for traps created very 
close to the disk surface.


\begin{figure}[!hhh]
 \includegraphics[width=0.45\textwidth]{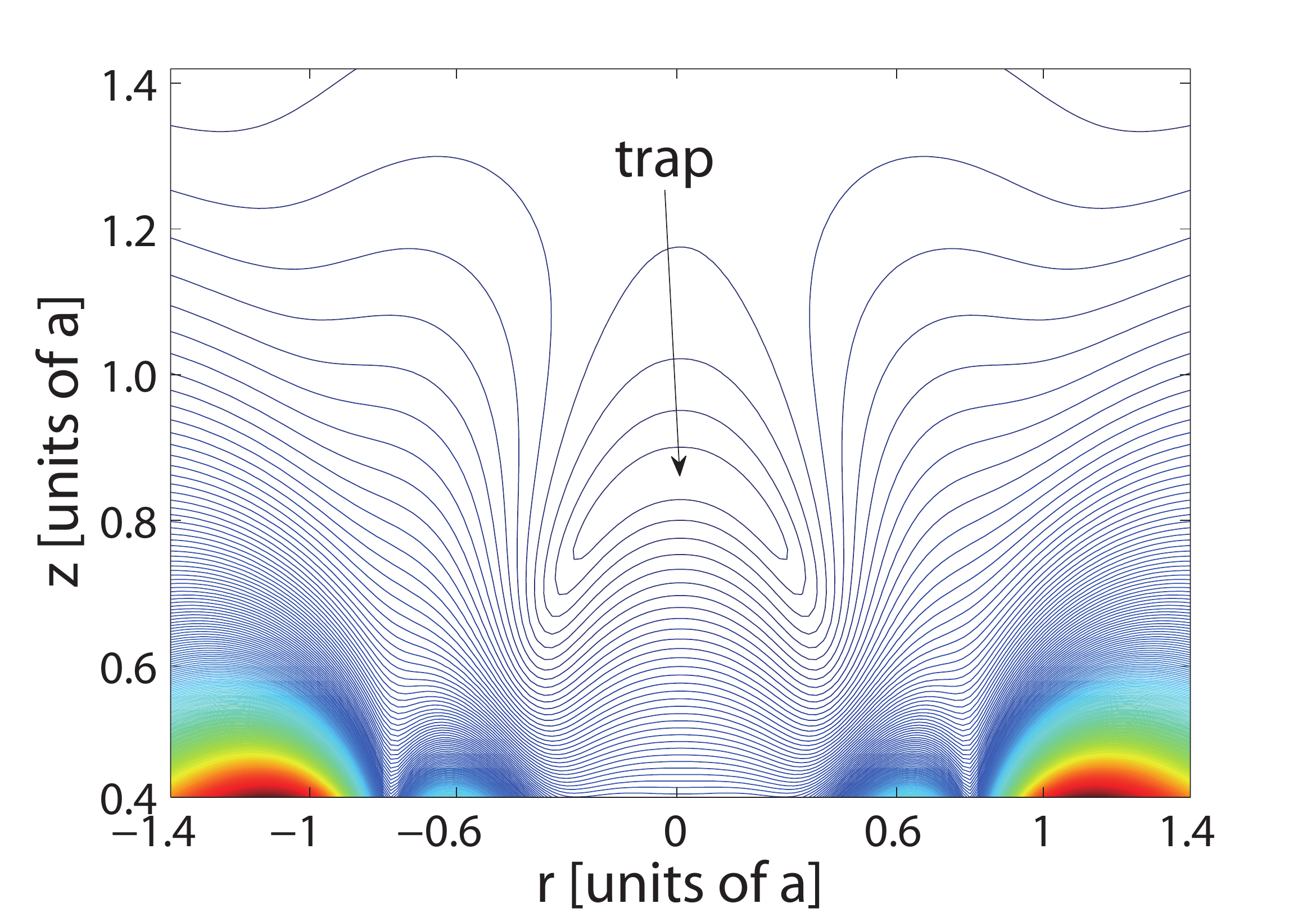}
    \caption{\label{fig:selftrap} (Color online) A self-sufficient trap above the disk center 
    formed by the vortex field created by two loading field pulses with 
    amplitudes $0.8 \mu_0 J_c$ and $-0.4 \mu_0 J_c$ respectively. The field variation per
    contour line is $9.5\times10^{-4} \mu_0 J_c$. }
 \end{figure}

The superconductor has a memory of the history of the external fields. 
Therefore it is possible to design a pulse sequence in which the role of the 
bias field is assumed by another induced supercurrent carried by the disk. 
Here we use two loading pulses with amplitudes of $0.8 \mu_0 J_c$ and $-0.4 
\mu_0 J_c$ respectively to write a vortex pattern in the disk, which results 
in a self-sufficient trap above the disk center, as shown in 
Fig.\ref{fig:selftrap}. Choosing typical experimental values for the critical current density of YBCO at liquid Nitrogen temperatures $j_c=1\mathrm{MA/cm^2}$ we calculate a self-sufficient trap depth of $35\mathrm{\mu K}$ for a disk with a diameter of $1\mathrm{mm}$. This trap has the feature of transforming into a ring 
trap by applying an additional perpendicular bias field $B_{bias}= 0.01 \mu_0 
J_c$.

While it is possible to split the self-sufficient trap into a ring, it does 
not necessarily imply a simple loading procedure with a high transfer 
efficiency from an external trap via the self-sufficient trap. A more 
straightforward way of producing a ring trap from an external quadrupole trap 
is to use a first magnetic field pulse of $1.2 \mu_0 J_c$ followed by a second 
pulse $-0.6 \mu_0 J_c$. Due to the nonlinear response of the material, this pulse sequence does not result in a 
self-sufficient trap, and instead a quadrupole trap can be formed far away 
from the chip surface by applying an additional bias field $B_{bias}$. By 
increasing the value of $B_{bias}$ this trap is brought closer to the chip 
surface where it eventually deforms into a ring trap. \linebreak
\begin{figure}[h]
\includegraphics[width=0.4\textwidth]{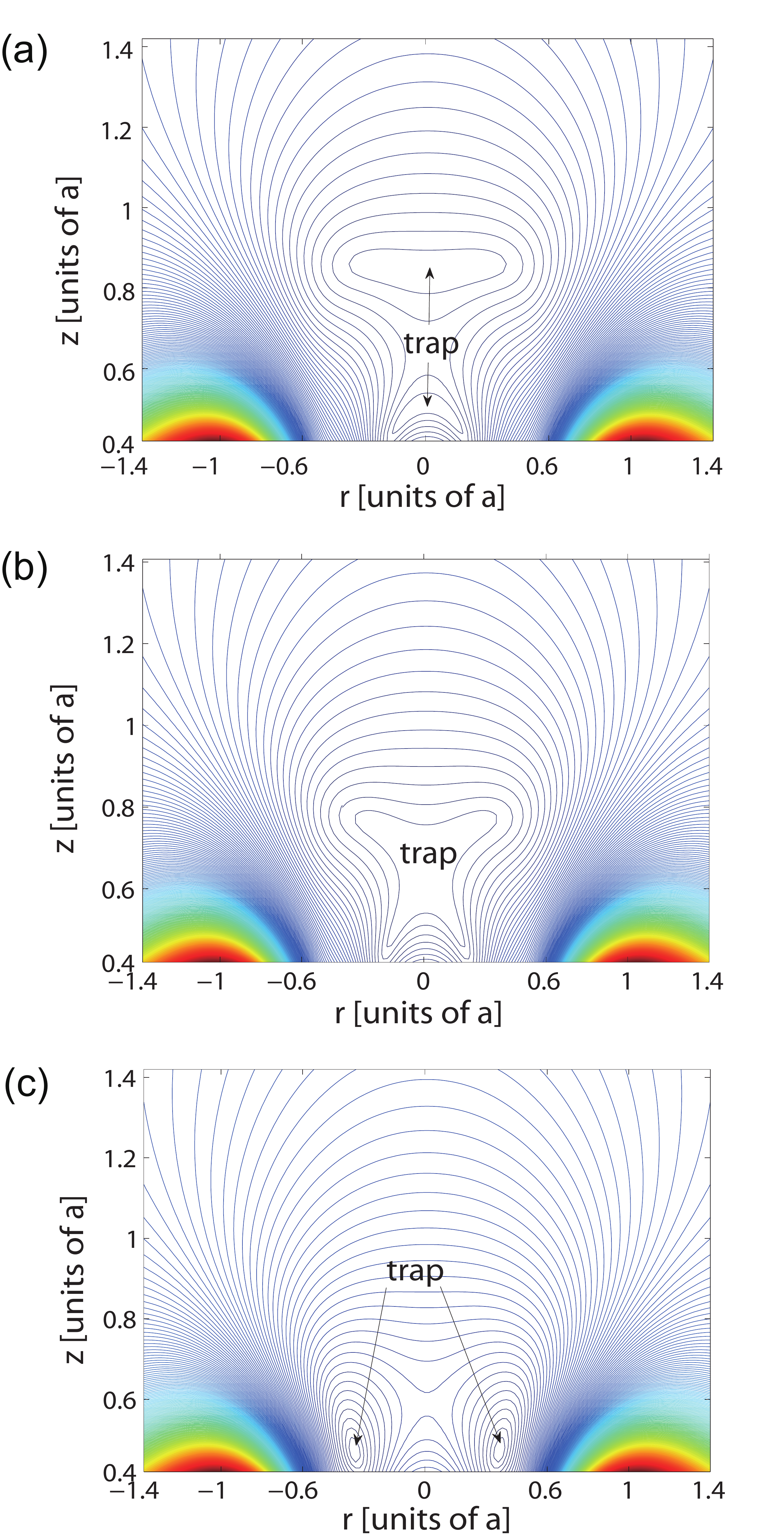}
    \caption{\label{fig:2trap} (Color online) Equipotential lines for an atom with nonzero 
    magnetic moment. The loading field pulses are $B_{ext}=0 \to 1.2 \mu_0 J_c 
    \to -0.6 \mu_0 J_c \to 0$. (a) With an additional bias field 
    $B_{bias}= 0.03 \mu_0 J_c$, two traps are obtained above the disk center. The field variation per
    contour line is $1.5\times10^{-3} \mu_0 J_c$. 
    (b) $B_{bias}= 0.035 \mu_0 J_c$, the two traps merge. The field variation per
    contour line is $1.54\times10^{-3} \mu_0 J_c$. (c) $B_{bias} = 0.06 
    \mu_0 J_c$, a ring trap is formed. The field variation per
    contour line is $1.2\times10^{-3} \mu_0 J_c$. 
    }
 \end{figure}
The perpendicular bias field lifts the field zero on the disk surface and 
lowers the quadrupole trap, producing two traps above the disk center, see 
Fig.\ref{fig:2trap}(a). Increasing the bias field brings the two traps closer 
together, until they merge at $B_{bias}= 0.03 \mu_0 J_c $, see 
Fig.\ref{fig:2trap}(b). Increasing $B_{bias}$ further will turn the trap into 
a ring trap, as shown in Fig.\ref{fig:2trap}(c). Continuing to increase 
$B_{bias}$ increases the radius of the ring trap until the trap is lowered to 
the surface.

\subsection{Traps generated by a superconducting ring}
\label{sec:ring}

\begin{figure}[!hhh]
 \includegraphics[width=0.4\textwidth]{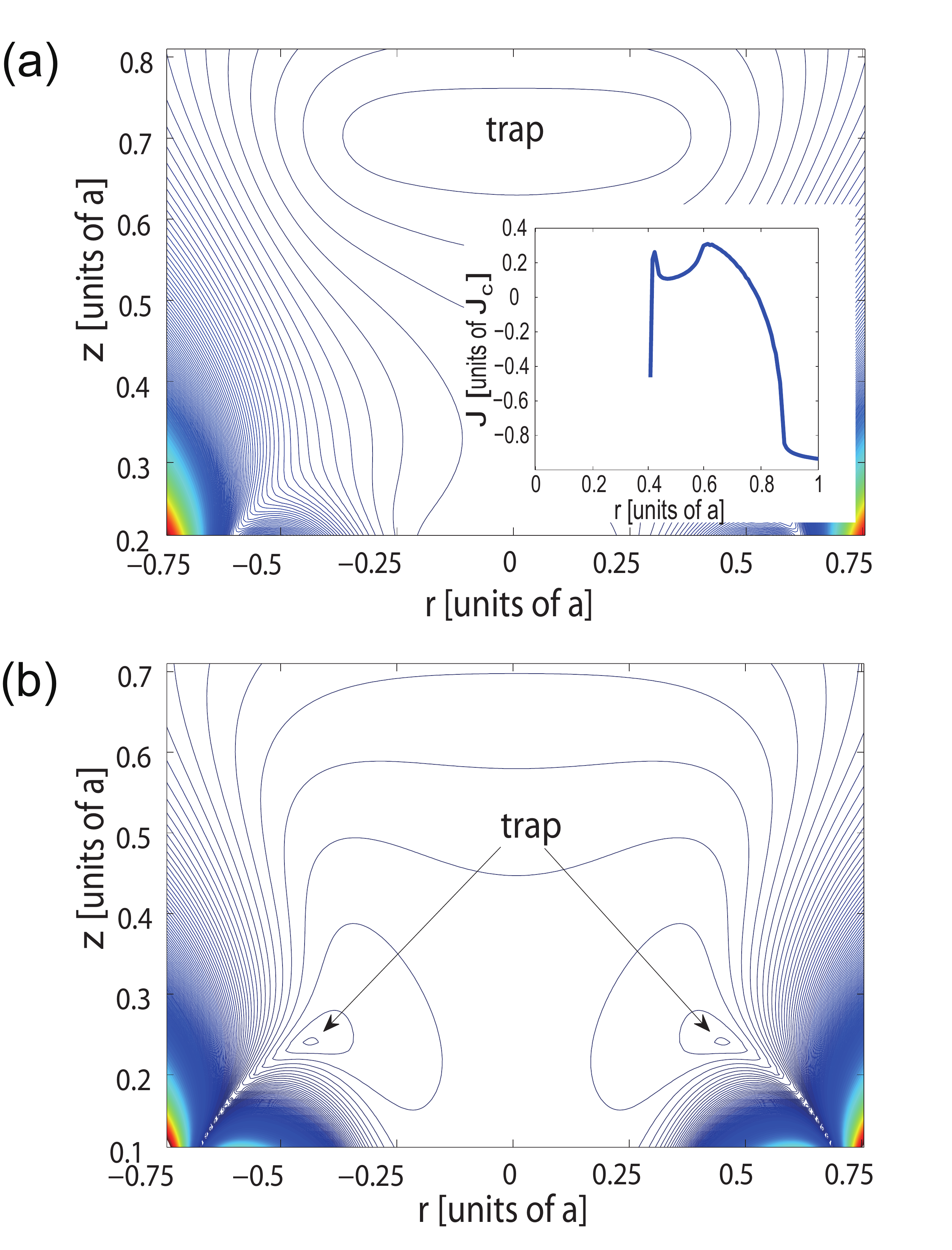}
   \caption{\label{fig:Brembias05} (Color online) (a) A quadrupole trap formed by a loading 
   field pulse $B_{ext}=0 \to 0.8\mu_0 J_c \to 0$ and a bias field 
   $B_{bias}= -0.028 \mu_0 J_c$. The field variation per
    contour line is $2\times10^{-3} \mu_0 J_c$. Inset: Distribution of the sheet current 
   density along the radial direction of the superconducting ring. (b) A 
   ring trap formed when increasing the bias field to $B_{bias}= -0.06 
   \mu_0 J_c$. The field variation per
    contour line is $5\times10^{-3} \mu_0 J_c$. }
 \end{figure}

In this section we consider an annular superconductor. The inset of 
Fig.\ref{fig:Brembias05}(a) shows the supercurrents carried by the vortices 
in the remanent state, when we apply a field pulse with a magnitude of 
$0.8\mu_0 J_c$ to a superconducting ring.\linebreak
\begin{figure}[!hhh]
 \includegraphics[width=0.35\textwidth]{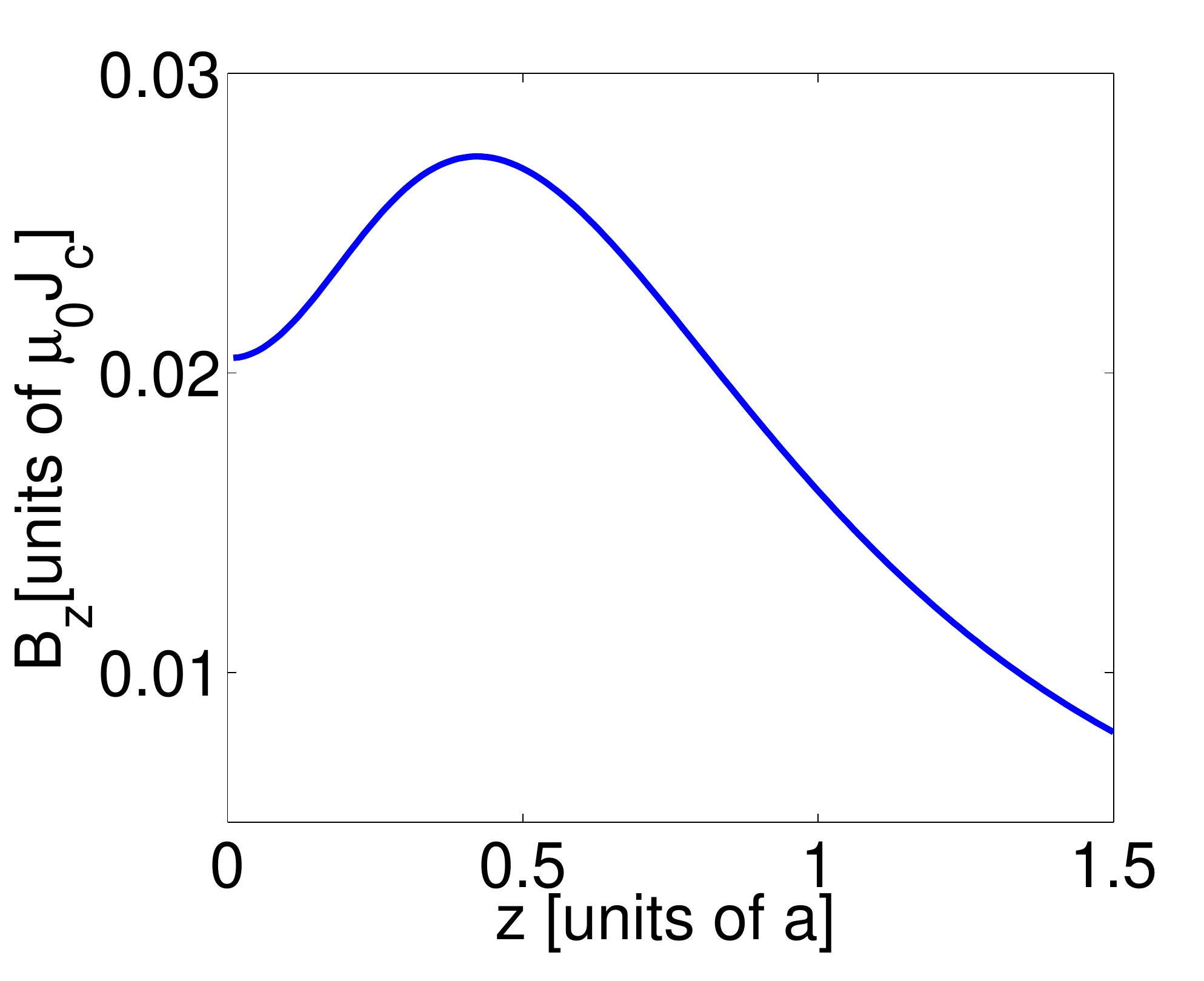}
   \caption{\label{fig:Jring} (Color online) $B_z(r=0,z)$ carried by the vortices at 
   different $z$ above the ring center. The loading field pulse is $B_{ext}=0 
   \to 0.8\mu_0 J_c \to 0$. }
 \end{figure} 
 \linebreak
 Similar to the superconducting disk, $B_r(r=0,z)$ is always zero by symmetry. 
 We plot the $z$ component of the magnetic field $B_z(r=0,z)$ generated by 
 these supercurrents at different height $z$ in Fig.\ref{fig:Jring}. 
 If an additional perpendicular field is applied to cancel $B_z(r=0,z=z_0)$, a 
 quadrupole trap can be formed at $(0, z_0)$. When $B_{bias}=-0.028\mu_0 J_c$ a 
 quadrupole trap is formed at $z_0 \sim 0.7a$,  as shown in 
 Fig.\ref{fig:Brembias05}(a). For small bias fields, increasing the magnitude 
 of $B_{bias}$ reduces the trap distance to the chip surface. When  
 $|B_{bias}|\sim 0.06 \mu_0 J_c$, the quadrupole trap transforms into a ring 
 trap, as shown in Fig.\ref{fig:Brembias05}(b). Further increasing the bias 
 field will enlarge the radius of the ring trap until the trap is lowered to 
 the surface. \linebreak
 \begin{figure}[!hhh]
 \includegraphics[width=0.45\textwidth]{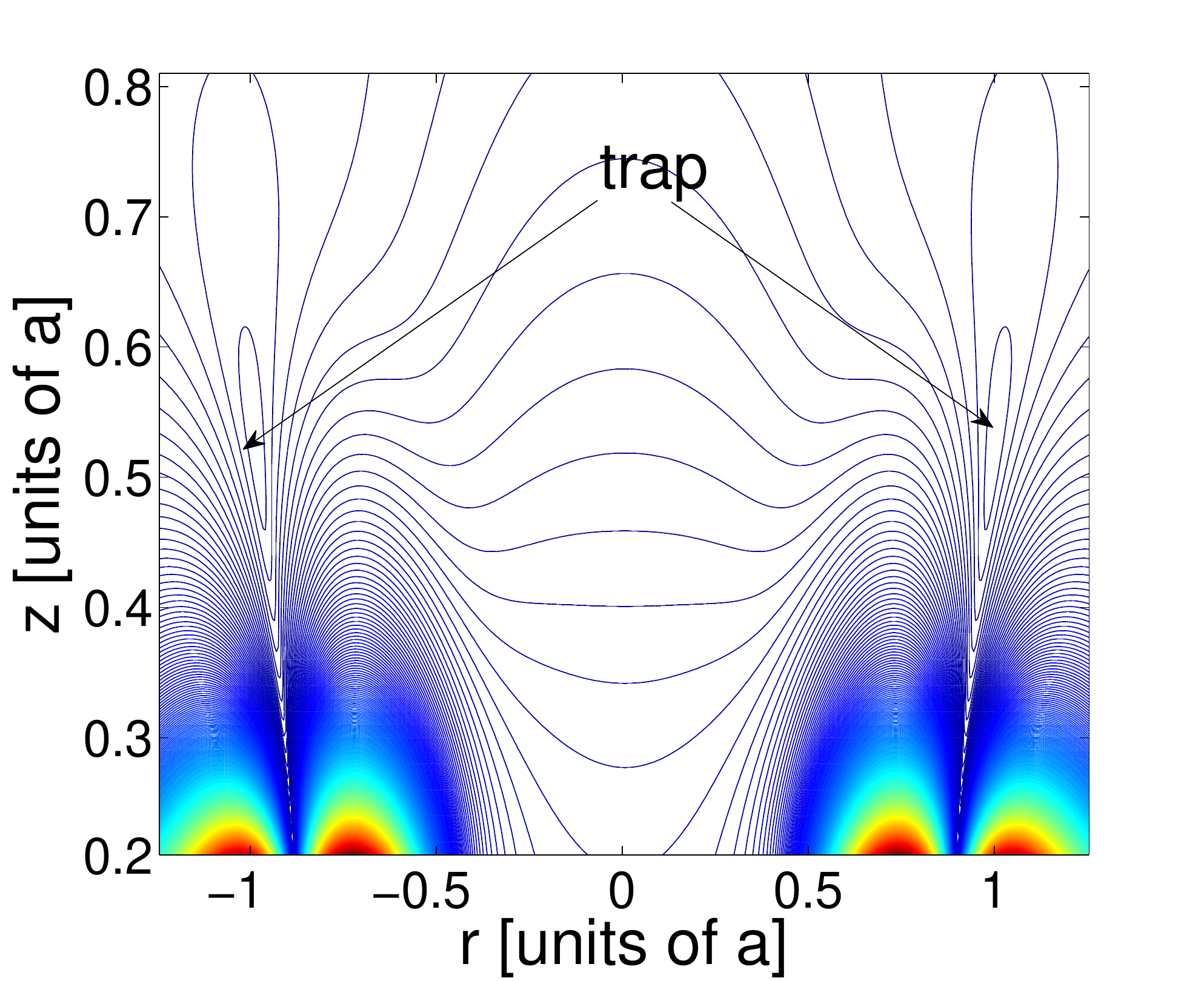}
   \caption{\label{fig:ringtrap2} (Color online) A ring trap formed by a loading field 
   sequence $B_{ext}=0 \to \mu_0 J_c \to -0.5 \mu_0 J_c \to 0$ and a bias 
   field $B_{bias} = 0.005\mu_0 J_c$. The double pulse sequence increases the 
   confinement of the atoms in the ring compared to the single pulse loading 
   sequence. The field variation per
    contour line is $1.8\times10^{-3} \mu_0 J_c$. }
 \end{figure} 
Alternatively, a ring trap may be formed by applying two field pulses and a 
bias field as indicated in Fig.\ref{fig:ringtrap2}. The first pulse has a 
magnitude of $\mu_0 J_c$, the second has a magnitude of $-0.5\mu_0 J_c$ and 
the bias field $B_{bias}=0.005\mu_0 J_c$. The ring trap of 
Fig.\ref{fig:ringtrap2} has a larger radius and distance to the surface as 
well as tighter confinement than the one shown in Fig.\ref{fig:Brembias05}(b).

 \section{Conclusion}
 \label{sec:con}

We have investigated and designed a variety of confining geometries for ultra 
cold neutral atoms based on type-II superconducting disk and ring structures. 
These geometries include quadrupole traps, a self-sufficient trap and ring 
traps. We have shown that a quadrupole trap above a superconducting disk and 
ring can be formed by applying a single magnetic field loading pulse and 
adding an external bias field. Using a double pulse sequence we are able to 
produce a self-sufficient trap above the disk center. We have developed a 
procedure for the formation of ring traps based on superconducting disk and 
ring structures. Furthermore we have shown an efficient loading procedure for 
these ring traps. 
Apart from methods outlined in this paper based on superconductors in the mixed state, it is possible to produce quadrupole traps with a ring-shaped superconductor in the Meissner state using flux trapped inside the ring and a bias field. By varying the trapped flux and bias field magnitudes ring traps can be produced. However, the limitations on the imprinted current distribution make it impossible to generate versatile magnetic field patterns \cite{BrandtRings}.
The key point of using type-II superconducting structures is that a single microstructured element is capable of generating tailored 
complex trap geometries for a variety of experimental implementations.
This allows us to control the geometry of confining potentials and their 
parameters without having to fabricate a new atom chip. In addition to this 
flexibility, superconductors have the intrinsic advantage of a low trap loss 
rate due to the reduced spin flips of the atoms close to the chip surface. The 
potential contribution to the spin flip rate by vortex motion in the 
superconductor has yet to be experimentally investigated.

An additional control over the trap geometries, complementary to the magnetic field pulses, can be obtained by dressing the atomic states with a radio-frequency field \cite{Schmied, Spreeuw}. This can help to lift the magnetic field zero intrinsically present in quadrupole traps, thus reducing the Majorana spin flip rate.
The low field noise in the proposed ring traps may make them an attractive 
candidate for the construction of a Sagnac-type interferometer \cite{Arnold}. 
A persistent flow of ultracold atoms, even an analog of the superconducting 
quantum interference device (SQUID), may be realized in these ring 
traps \cite{Ryu,Ramanathan}.
Taking into account all the intrinsic advantages discussed highlights the 
potential of magnetic traps for neutral atoms based on patterned vortex 
distributions in superconducting disks and rings.

\begin{acknowledgments}
We acknowledge financial support from Nanyang Technological University 
(grant no. WBS M58110036), A-Star (grant no. SERC 072 101 0035 and 
WBS R-144-000-189-305) and the Centre for Quantum Technologies, Singapore. 
We thank S.A. Cheong for fruitful discussion.
\end{acknowledgments}

\end{document}